\documentclass[utf8]{frontiersFPHY} 

\usepackage{url,hyperref,lineno,microtype,subcaption}
\usepackage[onehalfspacing]{setspace}

\usepackage{graphicx}
\usepackage{dcolumn}
\usepackage{bm}
\usepackage{longtable}
\usepackage{ulem}
\newcommand{\bea}{\begin{eqnarray}}
\newcommand{\eea}{\end{eqnarray}}
\newcommand{\beq}{\begin{equation}}
\newcommand{\eeq}{\end{equation}}

\newcommand{\simless}[0]{\mathbin{\lower 3pt\hbox
   {$\rlap{\raise 5pt\hbox{$\char'074$}}\mathchar"7218$}}}
\newcommand{\simgreat}[0]{\mathbin{\lower 3pt\hbox
   {$\rlap{\raise 5pt\hbox{$\char'076$}}\mathchar"7218$}}}

\newcommand{\figref}[1]{figure \ref{#1}}

\newcommand{\capfigref}[1]{Figure \ref{#1}}

\newcommand{\eqnref}[1]{equation (\ref{#1})}
\newcommand{\eqnrefs}[1]{equations (\ref{#1})}
\newcommand{\eqnrefbare}[1]{(\ref{#1})}
\newcommand{\capeqnref}[1]{Equation (\ref{#1})}

\def\keyFont{\fontsize{8}{11}\helveticabold }
\def\firstAuthorLast{Nadkarni-Ghosh {et~al.}} 
\def\Authors{Sharvari Nadkarni-Ghosh\,$^{1,*}$, Jayanta K. Bhattacharjee\,$^{2}$ }
\def\Address{$^{1}$Department of Physics, I.I.T. Kanpur, Kanpur 208016, India \\
$^{2}$Department of Physics, Harish-Chandra Research Institute, Jhusi, Allahabad 211019, India }
\def\corrAuthor{Sharvari Nadkarni-Ghosh}

\def\corrEmail{nsharvari@gmail.com, sharvari@iitk.ac.in}

\begin{document}
\onecolumn
\firstpage{1}

\title[Stability analysis of the plane Couette flow]{Stability analysis of fluid flows using Lagrangian Perturbation Theory (LPT): application to the plane Couette flow. }

\author[\firstAuthorLast ]{\Authors} 
\address{} 
\correspondance{} 

\extraAuth{}

\maketitle

\begin{abstract}
We present a new application of Lagrangian Perturbation Theory (LPT): the stability analysis of fluid flows. As a test case that demonstrates the framework we focus on the plane Couette flow. 
The incompressible Navier-Stokes equation is recast such that the particle position is the fundamental variable, expressed as a function of Lagrangian coordinates. The displacement due to the steady state flow is taken to be the zeroth order solution and the position is formally expanded in terms of a small parameter (generally, the strength of the initial perturbation). The resulting hierarchy of equations is solved analytically at first order.  We find that we recover the standard result in the Eulerian frame: the plane Couette flow is asymptotically stable for all Reynolds numbers. However, it is also well established that experiments contradict this prediction. In the Eulerian picture, one of the proposed explanations is the phenomenon of  `transient growth' which is related to the non-normal nature of the linear stability operator. The first order solution in the Lagrangian frame also shows this feature, albeit qualitatively.  As a first step, and for the purposes of analytic manipulation, we consider only linear stability of 2D perturbations but the framework presented is general and can be extended to higher orders, other flows and/or 3D perturbations. 

\tiny
 \keyFont{ \section{Keywords:} keyword, keyword, keyword, keyword, keyword, keyword, keyword, keyword} 
\end{abstract}

\section{Introduction}
Understanding the transition of fluid flows from the stable to turbulent regime is one of the central questions in the studies of turbulence. In usual linear stability analysis one formally expands the Navier-Stokes equation about a steady state flow assuming that the perturbations to the background flow are small. Study of linear stability of laminar flows, such as the plane Couette flow, commenced over a hundred years ago with the seminal work by Orr \cite{orr_1907} and Sommerfeld \cite{sommerfeld_1908}. The resulting Orr-Sommerfeld equation has been studied extensively over the decades (see Bayly {\it et al.} \cite{bayly_instability_1988} for a review). Purely analytic investigations were not conclusive \cite{joseph_eigenvalue_1968,case2_1960, lin_1961} and many efforts were devoted to obtaining a numerical solution of this equation. The basic idea was the same: expand the velocity in terms of an orthogonal set of basis functions and recast the system as an eigenvalue problem. But the analyses differed in their choices of basis functions, which resulted in different convergence rates.  The numerical results indicated that the plane Couette flow is linearly stable for all Reynolds numbers whereas the plane Poiseuille flow exhibits a transition to turbulence at $Re = 5772.22$ \cite{orszag_accurate_1971}. Further insight into these results was obtained by other combined numerical and analytical techniques \cite{hughes_variable_1972,davey_stability_1973,marcus_greens_1977}. 

However experimental results particularly for shear driven flows, show a discrepancy with linear predictions. For example, the plane Couette flow shows a transition to turbulence when none is expected while the plane Poiseuille transitions to turbulence at a Reynolds number much lower than the linear estimate \cite{tillmark_experiments_1992,daviaud_subcritical_1992,lemoult_experimental_2012}. Various approaches have been employed to explain this transition. One way is to computationally investigate the full non-linear Navier-Stokes equation as was done by Orszag and collaborators \cite{orszag_subcritical_1980,orszag_transition_1980} who showed that the energy growth in the system corresponds to a sub-critical bifurcation. Another is to look for finite amplitude equilibrium states near the transition and examine their stability against two and three dimensional perturbations \cite{cherhabili_finite-amplitude_1997}. A third approach is to understand the stability properties of the perturbed base flow \cite{barkley_stability_1999,dauchot_streamwise_1995}. Further developments in the 1990s
showed that the instability can be attributed to the `non-normality' of the linear stability operator \cite{schmid_studyofeigenvalues_1993,trefethen_pseudospectra_1997}. The eigenvectors of the linear operator are not orthogonal and this allows for the possibility of transient growth before the eventual asymptotic decay implied by the negative eigenvalues. The non-linear term can then amplify this growth \cite{butler_three-dimensional_1992,reddy_energy_1993, trefethen_hydrodynamic_1993}; see Grossmann \cite{grossmann_onset_2000} and Schmid \cite{schmid_nonmodal_2007} for recent reviews.

Majority of the analytical stability analysis has been carried out in the Eulerian frame. In this frame the velocity is the fundamental variable and is expressed as a function of a fixed Eulerian coordinate system (grid coordinates). On the other hand in the Lagrangian frame, the particle position is fundamental and is expressed as a function of a Lagrangian coordinate (usually the initial position) and time. Eulerian measurements are easier, whereas Lagrangian methods require sophisticated 3D particle tracking techniques (for example La Porta {\it et al} \cite{la_porta_fluid_2001,mordant_measurement_2001}). The choice of frame depends on the problem at hand. Much analytic and numerical work has been done in the Lagrangian frame in terms of analyzing particle statistics, predicting scaling laws, structure of correlation functions etc (see for example \cite{falkovich_2012} and references therein). In the context of geophysical flows, the Lagrangian picture has been used extensively to understand the backreaction effect of non-linear perturbations on the mean background flow following the  formulation by Andrews \& McIntyre \cite{andrews_1978}. Formal work regarding mathematical properties of the Lagrangian trajectories has been also performed \cite{zheligovsky_timeanalycity_2013} recently. However, linear stability analysis in this frame has been relatively rare.  

In this paper, we examine the stability of laminar flows using a perturbative scheme in the Lagrangian frame i.e., Lagrangian Perturbation Theory (LPT). As a simple test case, we focus only on 2D perturbations of the incompressible plane Couette flow but the formalism is general and can be extended to other flows. One of the motivations to use this method is that the Lagrangian derivative includes by definition the non-linear term $({\bf v} \cdot \nabla) {\bf v}$ and hence it ought to be able to better estimate the non-linear effect. Furthermore, a flow which is unstable in the Eulerian frame is also unstable in the Lagrangian frame. Thus, the investigation of Lagrangian stability can provide a independent confirmation of Eulerian stability. The main drawback of this scheme is that relies on the one-to-oneness of the  map between the Eulerian and the Lagrangian frame and fails when particles cross. It is not expected to model the turbulent regime where orbit crossing is likely to occur. 

LPT has been used in other branches of physics, most notably in cosmology, to model the growth of non-linear structure in the universe. 
The statistical theories of homogeneous and isotropic turbulence and the growth of non-linear cosmological large scale structure share many common features. The velocity field in the case of turbulence and the density field in the case of cosmology are both modeled as random fluctuations in a homogenous and isotropic background. In turbulence, the convective term in Navier-Stokes is the main source of non-linearity. Higher order velocity correlation functions are the main quantities of interest and one is interested in their scaling properties. In cosmology, the non-linearity arises both from the convective term in the Euler equation and gravity. Density correlations are of importance and they are used to constrain cosmological parameters. In the past, many perturbative techniques from the theory of turbulence have been successfully applied to analytically model cosmological structure. For example: Taruya and Hiramatsu \cite{taruya_closure_2008} use the Direct Interaction Approximation method of Kraichnan \cite{kraichnan_structure_1959,kida_lagrangian_1997,goto_direct-interaction_1998}  to address the `closure problem' of the hierarchy of moment equations. Crocce and Scoccimarro \cite{crocce_renormalized_2006} use the techniques discussed in Wyld \cite{wyld_formulation_1961} and L'vov \& Proccacia \cite{lvov_exact_1995} to formulate a renormalized perturbation theory. The adhesion approximation \cite{gurbatov_large-scale_1989}, used to deal with Lagrangian particle crossings  is essentially the model of 3D Burgers turbulence, see work by Frisch \& collaborators \cite{frisch_burgulence_2000, frisch_singularities_2001}, Gaite \cite{gaite_turbulence_2012}. 

We attempt to do the reverse: use a technique from the theory of large scale structure to understand the transition to turbulence. 
In this paper, we present the first step in applying LPT to the analysis of the incompressible Navier-Stokes equation. For simplicity we restrict to 2D perturbations and compute the first order solution of the scheme. The linear analysis using LPT analytically confirms the linear Eulerian stability result that the plane Couette flow is asymptotically stable at all Reynolds numbers. In addition it recovers the feature of transient growth, which in the Eulerian case is attributed to the non-normality of the linear stability operator. To the best of our knowledge a perturbative analysis in the Lagrangian frame has been performed in the past by Pierson \cite{pierson_perturbation_1962} in the context of geophysical flows. But our work differs from Pierson's because the order counting and flow geometries are different giving rise to a different set of equations and solutions. Pierson uses the Lagrangian particle labels as the zeroth order solution whereas we use the displacement due to the base flow as the zeroth order solution. The latter approach allows one to more easily track the dependence of the base flow making it easier to generalize.

The paper is organized as follows \S \ref{sec:basiceq} sets up the equations in the Lagrangian frame. \S \ref{sec:LPT} outlines the perturbative solution. 
The full solution needs to be computed numerically, but we show analytically that at late times the perturbations decay, confirming linear stability. \S \ref{sec:backtoEu} discusses the procedure to recover the Eulerian velocity from the Lagrangian velocity. \S \ref{sec:discussion} provides a discussion and summary.

\begin{figure}
\includegraphics[width=12cm]{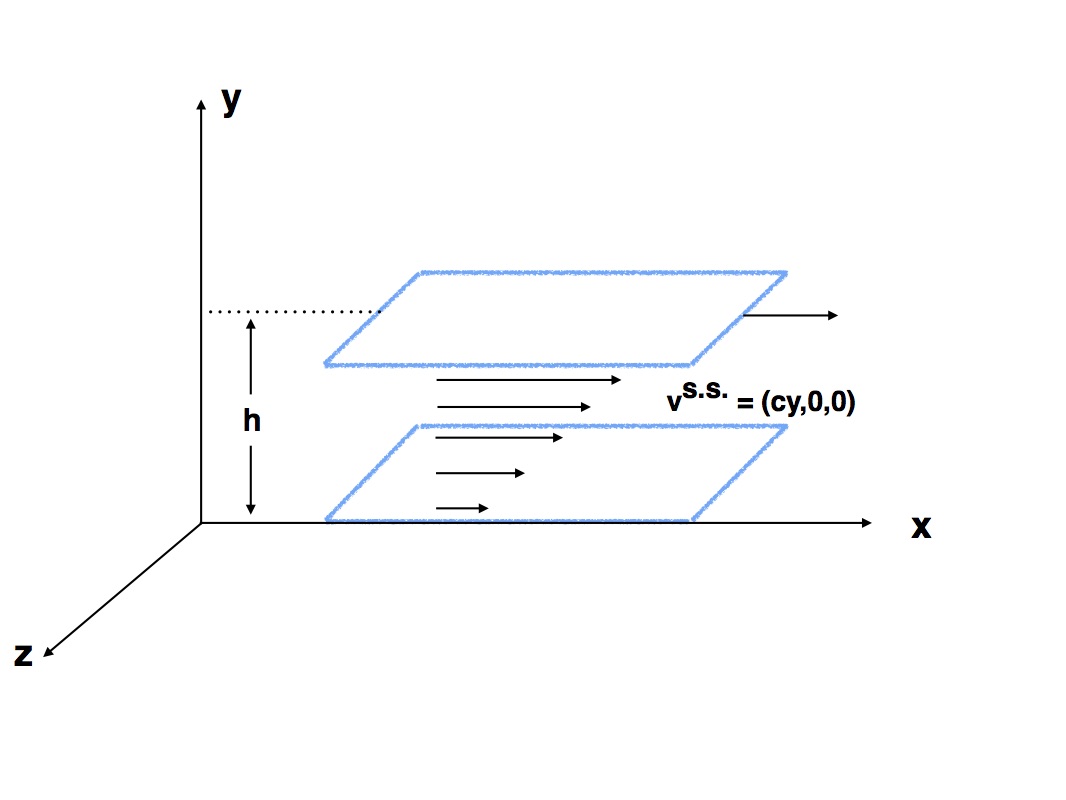}
\caption{Schematic representation of the 2D plane Couette flow. Two parallel plates are a distance $h$ apart; the steady laminar flow has a linear profile w.r.t. the $y$ coordinate i.e., ${\bf v}_{s.s.}=\{ cy,0,0\}$. For the semi-bounded case, $h \rightarrow \infty$.  }
\label{sketch}
\end{figure}

\section{Equations in the Lagrangian frame}
\label{sec:basiceq}
The incompressible fluid is described by the system of equations 
\bea
\label{curleq} \frac{d {\bf v}}{dt} &=&- \frac{\nabla P}{\rho} + \nu \nabla^2 {\bf v},\\
\label{diveq} \nabla \cdot {\bf v} &=& 0, 
\eea
where $d/dt$ is the usual convective derivative $\frac{d}{dt} = \frac{\partial}{\partial t} + ({\bf v} \cdot \nabla)$ and 
$P$, $\rho$ and $\nu$ denote the pressure, density and kinematic viscosity respectively. In this paper, we will focus only on laminar flows, in particular on the plane Couette flow which consists of two parallel plates moving with respect to each other with a steady state velocity ${\bf v}_{s.s.}$ (see \figref{sketch}. The velocity profile is invariant along the flow direction (defined to be the $x$-axis) and varies only in the direction perpendicular to the flow (defined to be the $y$-axis).  Let ${\bf r} = \{x,y,z\}$ denote the physical position of a fluid element. We will restrict to 2D perturbations and hence the fluid displacements are confined only to the $xy$-plane. The system given by \eqnrefs{curleq} and \eqnrefbare{diveq} is Eulerian:  the velocity is the fundamental quantity and is usually solved in terms of the fixed coordinate system (grid coordinates) i.e. ${\bf v} = {\bf v}_E({\bf r})$, where the subscript `$E$' denotes Eulerian. 
On the other hand, in the Lagrangian framework, the particle position is the fundamental quantity and is usually solved in terms of some fixed Lagrangian coordinate and time. We choose the Lagrangian coordinate ${\bf R}= \{X,Y,Z\}$ to be the physical position at the initial time ($t=0$). Thus, 
\beq 
{\bf R} = {\bf r}({\bf R}, 0).
\label{lagcoordef}
\eeq
The physical position at any later time is 
\beq 
{\bf r} \equiv {\bf r}({\bf R}, t) 
\eeq
and the corresponding Lagrangian velocity is 
\beq
{\bf v}_L = {\dot {\bf r}}({\bf R}, t), 
\eeq
where the subscript `$L$' denotes Lagrangian and `dot' denotes the time derivative $d/dt$. Note that in the Lagrangian frame the spatial variable ${\bf R}$ does not change with time. Thus, $d/dt$  is not represented in terms of a partial time derivative and the convective term as is done in the Eulerian frame;  instead it is the total derivative and is also denoted as $D/Dt$ in the literature. 

Taking the curl of \eqnref{curleq} and making ${\bf r}$ the fundamental variable, the system of \eqnrefs{curleq} and \eqnrefbare{diveq} can be equivalently expressed as 
\bea
\label{mainT} \nabla_r \times {\ddot {\bf r}} &=& \nu \nabla_r^2 (\nabla_r \times {\dot {\bf r}}),\\
\label{mainL} \nabla_r \cdot {\dot {\bf r}} &=& 0.
\eea
The spatial derivatives in the above equation are with respect to the physical variable ${\bf r}$. These have to be transformed to derivatives with respect to the Lagrangian coordinate (see appendix \ref{mathtrans}). This yields the system 
 \bea
\nonumber\epsilon_{ijk} \epsilon_{jmn} \epsilon_{lm'n'}  {\ddot r}_{k,l}r_{m,m'} r_{n,n'} &=& \nu \epsilon_{ijk} \epsilon_{jfg} \epsilon_{mf'g'} \epsilon_{lde} \epsilon_{qd'e'} \epsilon_{lab} \epsilon_{pa'b'} r_{d,d'} r_{e,e'}  \\ 
&& \left\{ \frac{1}{2J}  r_{a,a'} r_{b,b'} \left( \frac{1}{2J}  {\dot r}_{k,m} r_{f,f'} r_{g,g'}  \right)_{,p} \right\}_{,q} 
 \label{mainT2}\\
  \epsilon_{ilm} \epsilon_{jl'm'} {\dot r}_{i,j} r_{l,l'} r_{m,m'} &=&0,
\label{mainL2} 
   \eea
where the commas denote derivatives with respect to the Lagrangian spatial coordinate and 
 \beq
  J = {\rm Det} \left(\frac{\partial r_i}{\partial R_j} \right) = \frac{1}{6} \epsilon_{ijk}\epsilon_{lmn}r_{i,l}r_{j,m} r_{k,n}
 \eeq 
 is the determinant of the Jacobian of the transformation between the Eulerian and Lagrangian coordinates. 
Note that the time derivative commutes with the spatial Lagrangian differentiation i.e., `dots' and `commas' commute.

\section{Lagrangian perturbation theory (LPT)}
\label{sec:LPT}
In the Lagrangian perturbative scheme, ${\bf r}$ is formally expanded as 
\beq 
{\bf r}({\bf R}, t) = \sum_{n=0}^\infty {\bf p}^{(n)}({\bf R}, t) \Delta^n
\label{EXP}
\eeq
where ${\bf p}^{(n)}$ is the $n$-th order term and $\Delta$ is just a formal book-keeping parameter related to the strength of the initial velocity perturbation. The definition of the Lagrangian coordinate (\eqnref{lagcoordef}) is used to set the initial values of the displacement vectors at each order. We choose 
\bea 
{\bf p}^{(0)}({\bf R}, 0) &=& {\bf R},
\label{initp0}\\
 {\bf p}^{(n)} ({\bf R}, 0) &=& 0 \; \; \; \forall \; n>0.
 \label{initpn}
 \eea 
\subsection{Zeroth order}
The background steady state solution for the plane Couette flow is given by ${\bf v}_E^{s.s.} = \{cy,0,0\}$. By definition, $y =Y$ at the initial time. Then the {\it initial} velocity in the Lagrangian frame is ${\bf v}_L({\bf R}, 0) = \{c Y,0,0\}$. Thus the particle at initial position ${\bf R}$ at $t=0$ is at ${\bf R} + {\bf v}_L({\bf R}, 0) t$ after time $t$. We take this to be the zeroth order solution for the position vector i.e. 
\beq 
{\bf p}^{(0)}({\bf R}, t) = {\bf R} + {\bf v}_L({\bf R},0) t.
\label{zeroth}
\eeq
In the Eulerian framework, stability of the flow is examined by perturbing around the steady state velocity in Eulerian coordinates. The zeroth order solution in that case is always an exact solution of the incompressible Navier-Stokes system. It is necessary to check that the transformation to Lagrangian coordinates preserves this property of the background solution. This is shown in appendix \ref{zeroorder}.

\subsection{First order equations}
Substitute the ansatz of \eqnref{EXP} in the system of \eqnrefs{mainT2} and \eqnrefbare{mainL2} and keep terms up to first order. 
Using the zeroth order solution given by \eqnref{zeroth} (see appendix \ref{perttheo} for details), the equation for the first order displacement is 

\begin{flushleft}   
\beq
\label{diveqfin}\nabla_R \cdot {\dot {\bf p}^{(1)}} = c \left(p^{(1)}_{Y,X} + t {\dot p}^{(1)}_{Y,X}\right)
\eeq
\beq
\label{curleqfin} {\ddot p}^{(1)}_{Y,X} - {\ddot p}^{(1)}_{X,Y} +  ct {\ddot p}^{(1)}_{X,X} = \nu \left( \nabla_R^2 -2 ct \frac{\partial^2}{\partial X \partial Y } +  c^2 t^2 \frac{\partial}{\partial X^2}  \right)  \left({\dot p}^{(1)}_{Y,X} - {\dot p}^{(1)}_{X,Y} +ct {\dot p}_{X,X} ^{(1)} -c p^{(1)}_{X,X}\right).
\eeq
\end{flushleft}
Here $p^{(1)}_{Y,X}$ denotes the derivative of the $Y$ component of ${\bf p}^{(1)}$ w.r.t. $X$ (partial spatial derivative).    In the usual Eulerian perturbation theory the perturbed velocity also remains divergence-free at first (and higher) orders; this condition arises because the flow is incompressible. This condition translated into the Lagrangian frame at first order gives \eqnref{diveqfin}. Note that the divergence of the perturbed velocity in the Lagrangian frame is not zero at any order; the non-zero terms arise from the transformation between the Eulerian and Lagrangian coordinate. In order to satisfy \eqnref{diveqfin}, we assume ${\bf p}^{(1)}$ to have the form  
\beq
{\bf p}^{(1)} (X,Y, t) = \left\{ \frac{\partial \psi}{\partial Y}-ct \frac{\partial \psi}{\partial X}, -\frac{\partial \psi}{\partial X},0\right\},
\label{p1form}
\eeq
where, $\psi$ is a scalar function of Lagrangian coordinates and time i.e., $\psi \equiv \psi(X,Y,t)$.
This gives 
\beq
{\dot {\bf p}}^{(1)} (X,Y,t) = \left\{ \frac{\partial {\dot \psi}}{\partial Y}- c\frac{\partial \psi}{\partial X} -ct \frac{\partial {\dot \psi} }{\partial X}  , -\frac{\partial {\dot \psi}}{\partial X},0\right\}.
\label{p1dotform}
\eeq
$\psi$ is analogous to a stream-function, but not the same as one encountered in the usual Orr-Sommerfeld analysis. Substituting the form of \eqnref{p1form} in \eqnref{curleqfin} and simplifying gives
\begin{flushleft}
\beq
  \frac{d}{dt} \left[(1+c^2t^2)\frac{\partial^2}{\partial X^2} + \frac{\partial^2}{\partial Y^2} -2ct \frac{\partial^2}{\partial X \partial Y}    \right] {\dot \psi} = \nu \left[(1+c^2t^2)\frac{\partial^2}{\partial X^2} + \frac{\partial^2}{\partial Y^2} -2ct \frac{\partial^2}{\partial X \partial Y}    \right]^2 {\dot \psi}.
\label{psieq}
\eeq
\end{flushleft}
This system can be recast as 
 \bea
{\mathcal A} {\dot \psi}  &=& \phi,
 \label{simpleeq1} \\
{\dot \phi } &=& \nu \mathcal A \phi, 
 \label{simpleeq2}
\eea
where the operator ${\mathcal A}$ 
 \beq
 \mathcal{A} = (1+c^2t^2)\frac{\partial^2}{\partial X^2} + \frac{\partial^2}{\partial Y^2} -2ct \frac{\partial^2}{\partial X \partial Y}.
 \eeq
The solution is obtained by first solving \eqnref{simpleeq2} for $\phi$ and then solving \eqnref{simpleeq1} for ${\dot \psi}$. Integrate to get $\psi$. 

\subsection{Initial and boundary conditions}
The symmetry of the underlying flow implies periodic boundary conditions along the $X$-axis for solving both $\phi$ and ${\dot \psi}$. We assume that the $X$-dependent part of the solution can be separated from the rest and represent it by a Fourier series expansion. With this ansatz, the net system defined by \eqnrefs{simpleeq1} and \eqnrefbare{simpleeq2} is second order in time and fourth order in the spatial variable $Y$. Accordingly two temporal initial conditions and four boundary conditions (two for ${\dot \psi}$ and two for $\phi$) are needed. 

 The perturbation velocity profile at $t=0$ is specified initially. In numerical simulations this is done by exciting specific modes or specifying an initial power spectrum of the perturbation. Let ${\bf v}_0(X,Y)$ formally denote this initial perturbation. 
\beq 
 {\dot {\bf p}}^{(1)}(X,Y, t=0) = {\bf v}_0(X,Y).
\label{initpdot}
 \eeq 
 The definition of the Lagrangian coordinate provides the other initial condition. \capeqnref{initpn} for $n=1$ is 
\beq 
 {\bf p}^{(1)}(X,Y, t=0) =0 \; \; \; \forall \; X,Y.
 \label{initp}
\eeq 
The boundary conditions are more involved. The symmetry of the underlying flow implies periodic boundary conditions along the $X$-axis for solving both $\phi$ and ${\dot \psi}$. The no slip condition imposed on the wall at $Y=0$ means 
\beq 
 {\dot {\bf p}}^{(1)}(X,Y=0,t) =0 \; \; \; \forall \; t.
 \label{boundpdot}
\eeq 
These are two conditions corresponding to the $X$ and $Y$ coordinate. Note that ${\dot {\bf p}}$ depends explicitly on $\psi$ and ${\dot \psi}$ but only indirectly on $\phi$. For simplicity we will assume that the flow is semi-bounded i.e., wall is placed only at $Y=0$. This allows us to 
relate \eqnref{boundpdot} to conditions on ${\dot \psi}$. 
Using the definition of ${\dot {\bf p}}^{(1)}$ from \eqnref{p1dotform} in \eqnref{boundpdot} gives
\bea
\label{bc0}\left. \frac{\partial {\dot \psi}}{\partial Y}\right|_{Y=0}  &=& \left. c\frac{\partial \psi}{\partial X}\right|_{Y=0} \; \; \; \forall \; t,\\
\left. \frac{\partial {\dot \psi}}{\partial X}\right|_{Y=0}  &=& 0 \; \; \; \; \; \;\;\;\;\; \forall \; t.
\label{bc1} 
\eea
Using \eqnref{bc1} and using the fact that the time derivative commutes with the spatial derivative gives $ \frac{d}{dt} \left(\left. \frac{\partial {\psi}}{\partial X}\right|_{Y=0}\right) = 0$, at all times $t$. Evaluating the $Y$-coordinate of \eqnref{initp} at $Y=0$ using \eqnref{p1form} gives $\left. \frac{\partial {\psi}}{\partial X}\right|_{Y=0}  = 0$ at the initial time $t=0$. Since the time derivative is zero, $\left. \frac{\partial {\psi}}{\partial X}\right|_{Y=0}$ stays zero at all times. Thus the r.h.s. of \eqnref{bc0} is zero at all times and it follows that 
\beq
\left. \frac{\partial {\dot \psi}}{\partial Y}\right|_{Y=0} = 0  \; \; \; \; \; \; \forall \; t.
\label{bc2}
\eeq
We assume periodic b.c. (Fourier decomposition) along the $Y$ axis for $\phi$. \capeqnref{bc1} and \eqnrefbare{bc2} provide the boundary conditions for ${\dot \psi}$.

\subsection{First order solution}
We now solve for $\phi$ and $\psi$ subject to the above conditions. \capeqnref{simpleeq2} for $\phi$ has spatial and temporal derivatives appearing on different sides of the equation and hence is separable. Following the arguments in the earlier section, the ansatz for $\phi$ is 
\beq 
\phi (X,Y,t) = \sum_{k_x, k_y} \tilde{\phi}(k_x, k_y) e^{i k_x X}  e^{i k_y Y} f(t). 
\label{phiform}
\eeq
Substituting in \eqnref{simpleeq2}, gives 
\beq 
\frac{d f}{dt} = \nu \left(-k_x^2 (1+c^2t^2) -k_y^2 + 2 ct k_x k_y \right) f. 
\eeq
The solution is 
\beq 
 f(t) = f(0) e^{\nu \left(-k^2 t  + k_x k_y c t^2 - \frac{k_x^2 c^2 t^3}{3}\right)}
\label{foft}
\eeq
where $k^2 = k_x^2 + k_y^2$ and $f(0)$ is the integration constant to be set later. The solution for $\phi$ is 
\beq
\phi(X,Y,t)  = f(0) \sum_{k_x, k_y} \tilde{\phi}(k_x, k_y) e^{i k_x X}  e^{i k_y Y}  e^{\nu \left(-k^2 t  + k_x k_y c t^2 - \frac{k_x^2 c^2 t^3}{3}\right)}.
\label{phisoln}
\eeq
One can now solve \eqnref{simpleeq1} for ${\dot \psi}$. From the structure of the equation one can assume that the purely temporal functions are the same for both $\phi$ and ${\dot \psi}$. Any additional time dependence introduced via the operator ${\mathcal A}$ is necessarily also a function of $Y$. This gives the ansatz  
\beq 
{\dot \psi}(X,Y,t) =  \sum_{k_x, k_y} \tilde{\phi}(k_x, k_y) e^{ik_x X} g_{k_y}(Y,t) f(t). 
\eeq
The subscript $k_y$ denotes the dependence of $g$ on the parameter $k_y$. We will drop it in subsequent evaluations. Substituting in \eqnref{simpleeq1} and using \eqnref{phisoln} gives 
\beq
-k_x^2 (1+c^2 t^2) g(Y,t) -2 ik_x ct g'(Y,t) + g''(Y,t) = e^{ik_yY}, 
\eeq 
where the primes denote differentiation w.r.t $Y$. 
The solution for $g(Y,t)$ can be split into a homogenous part and a particular solution: $g(Y,t) = g_{homo.}(Y,t) + g_{part.}(Y,t)$ where 
\beq 
g_{homo.}(Y,t) = C_1(t) e^{(-k_x +  k_x ict)Y} + C_2(t) e^{(k_x + k_x ict)Y}  
\eeq
and 
\beq 
g_{part.}(Y,t) = C_3(t) e^{ik_y Y}, 
\eeq
where 
\beq 
C_{3}^{-1}(t) =  -k_x^2(1+c^2t^2) -k_y^2 + 2k_x k_y ct.
\label{C3}
\eeq
Satisfying the boundary conditions represented by \eqnrefs{bc1} and \eqnrefbare{bc2} fixes $C_1(t)$ and $C_2(t)$.  
\beq 
C_1^{-1}(t) = 2 k_x (kx (1-ict) + ik_y), \; \; \; \; C_2^{-1}(t) = 2 k_x (kx (1+ ict) - ik_y). 
\eeq
The net solution for $g$ is 
\beq 
g(Y,t)   = C_1(t) e^{(-k_x +  ict k_x )Y} + C_2(t) e^{(k_x + ict  k_x )Y} + C_3(t) e^{ik_y Y}.
\eeq
$\psi$ is obtained by integrating ${\dot \psi}$ w.r.t. time:
\beq 
\psi =  \sum_{k_x,k_y} \tilde{\phi}(k_x, k_y) e^{ik_x X} \int g(Y,t') f(t')dt'  + C(X,Y), \\
\label{psi}
\eeq
where $C(X,Y)$ is the constant of integration. This can be re-written as 
\beq
\psi =  \sum_{k_x, k_y} f(0) \tilde{\phi}(k_x, k_y) e^{ik_x X} h(Y,t) + C(X,Y),
\eeq where $h(Y,t) = \int g(Y,t) f(t)/f(0) dt$. The initial condition \eqnref{initp} is satisfied if $\psi=0$ at $t=0$. This fixes $C(X, Y) = -f(0) \tilde{\phi} (k_x, k_y) e^{ik_x X} h(Y,t=0)$ giving 
\beq 
\psi = \sum_{k_x, k_y} f(0) \tilde{\phi}(k_x, k_y) e^{ik_x X} [h(Y,t) - h(Y,t=0)].
\eeq
The only quantities remaining to be determined are $f(0)$ and $\tilde{\phi}(k_x, k_y)$. This is set by the initial velocity perturbation. The individual components $\phi(k_x, k_y)$ and characterize its shape and $f(0)$ sets the overall initial amplitude. 

{\it Late time behaviour}:  The above prescription completely specifies the initial conditions and the numerical solution for $\psi$ can be computed at any later time.  
Simplified analytic expressions can be obtained at late times. Note that the terms arising from the homogenous part of the solution are damped and oscillatory. If the integration is over a sufficiently large time interval $t >> (c k_x)^{-1}$ they integrate to zero leaving  
\beq 
\psi \sim  e^{ik_x X} e^{ik_y Y} \int C_3(t') f(t') dt'
\eeq 
Substituting for $C_3$ and $f$ from \eqnrefs{foft} and \eqnrefbare{C3} gives 
\bea
\psi &\sim& {\tilde \psi}(k_x, k_y) f(0) e^{ik_x X}e^{ik_y Y} \int - \frac{e^{-\frac{\nu k_x^2 c^2 t'^3}{3}}}{k_x^2 c^2 t'^2} dt' \\
&\sim & \frac{f(0)}{k_x^2 c^2 t} \left[e^{-\frac{\nu k_x^2 c^2 t^3}{3}} + t \left( \frac{\nu k_x^2 c^2 }{3}\right)^{1/3}  \gamma\left(\frac{2}{3},  \frac{\nu k_x^2 c^2 t^3 }{3}\right) \right] + C(X,Y)
\eea
where $\gamma(s,z) = \int_o^z e^{-z'} z'^{s-1} dz' $ is the lower incomplete gamma function. 

Thus $\psi$ and hence the velocity perturbation evolves as $\sim e^{-\frac{\nu k_x^2 c^2 t^3}{3}}$ at late times and the flow is linearly stable in the Lagrangian frame. This late time behaviour is in exact agreement with that of Press \& Marcus \cite{marcus_greens_1977} which was obtained using symmetry arguments for the unbounded Couette flow \footnote{We have chosen dimensional units since for the semi-bounded flow, there is no inherent length scale to define the Reynolds number. Thus, ${\bf p}$ is a displacement and $\psi$ has dimensions of $m^2$; $g \sim m^2; f \sim t^{-1}$. }. Once $\psi$ is known, the full first order displacement and velocity can be computed from \eqnrefs{p1form} and \eqnrefbare{p1dotform}. 

In the above analysis we used Fourier basis functions for both the $X$ and $Y$ axis in solving for $\phi$. This allowed us to solve for $\phi$ and $\psi$ in succession. If the flow is bounded at some finite $Y$ then the boundary conditions are more complicated. One has to choose appropriate basis functions which will satisfy them. Although we do not provide a complete numerical solution for this case, we briefly sketch its form in appendix \ref{app:bounded}. 
We note that the temporal function $f(t)$ has a similar exponential form; velocity perturbation evolves as $\sim e^{-\frac{\nu k_x^2 c^2 t^3}{3}}$ at late times. Defining the Reynolds number for the bounded flow of height $h$ as $Re = v^{s.s.}(h) \cdot h /\nu = ch^2/\nu$, the perturbation evolves as $\sim \exp {(-\frac{h^2 k_x^2 c^3 t^3}{3 Re}})$.

\subsection{Discussion}
\label{sec:discussion}
The main temporal dependence of the Lagrangian solution is given by the exponential term in \eqnref{foft}. All three terms in the exponent arise from the viscous component as is indicated by the factor $\nu$ multiplying them. The linear term $-\nu k^2 t$ in the exponent also arises in standard Eulerian linear perturbation theory, but quadratic and cubic terms are new in the Lagrangian frame. It is also interesting to note that the solution for a small time interval can grow as $e^{\nu k_x k_y c t^2}$ before eventually falling off as $e^{-\nu k_x^2 c^2t^3/3}$ at late times. For sufficiently large $k_y$, this term can dominate the dynamics. This hints at the phenomenon of transient growth which is believed to be responsible for the instabilities observed in shear flow experiments. In Eulerian perturbation theory this transient growth has been attributed to the fact that the linear stability operator is non-normal. The stability in these cases is not governed just by the spectrum, but by its pseudospectrum \cite{schmid_studyofeigenvalues_1993,trefethen_pseudospectra_1997}. However,  the linear transient growth explanation has also found critics \cite{waleffe_1995}. In particular, the effect of the non-linearity on the mean background flow plays an important role in the transition to turbulence. Recently, Fukumoto {\it et al} \cite{fukumoto_2010} used Lagrangian approach to study the weakly non-linear stability in the case of elliptical flows. Understanding the transient growth and weakly non-linear regime for the plane Couette flow using LPT is left for future work.

Stability analyses in the Eulerian and Lagrangian frame differ in one fundamental aspect. In the former, non-linear convective terms such as $({\bf v} \cdot \nabla)  {\bf v}$, which are second order in the velocity perturbation are ignored whereas in the Lagrangian frame, they get absorbed into the time derivative operator. This suggests a potential advantage of the Lagrangian frame over the Eulerian frame. However, other than the fact that we get exponents that are quadratic and cubic in $t$, we do not get qualitatively different results at linear order in the two frames. In appendix \ref{app:trans} we estimate the effect of the convective term in Eulerian perturbation theory using an interative procedure. It shows qualitatively a similar behaviour as compared to the first order Lagrangian solution, but there is no quantitative agreement nor does it provide any additional quantitative insights into the nature of transient growth. Finally, we must remind ourselves that there is never a clear matching between orders in the Eulerian and Lagrangian frame. A more quantitative comparison of the linear LPT solution to that obtained in the Eulerian picture can be done only when the Lagrangian to Eulerian map is inverted as is described in the next section (see \S \ref{sec:backtoEu}). This requires further numerical work and is beyond the scope of this paper.

\section{Transforming back to the Eulerian frame}
\label{sec:backtoEu}
The LPT scheme outlined in the previous section solves \eqnrefs{mainT} and \eqnrefbare{mainL} for the variable ${\bf r}$.
However, the original system whose solution we seek is given by \eqnrefs{curleq} and \eqnrefbare{diveq}. \capeqnref{mainT} is obtained by taking spatial derivatives of \eqnref{curleq} and hence the LPT solution is insensitive to any spatially homogenous time dependent transformation $\Delta {\bf r}(t)$.  In recent work, Nadkarni-Ghosh \& Chernoff \cite{nadkarni-ghosh_modelling_2013} showed that convergence properties of the perturbative solution crucially depend on fixing this degree of freedom, although the work was in the context of a different physical system, namely dark matter fluid gravitating an expanding universe. Since our aim in this paper is to merely examine the stability, we do not explicitly calculate the exact form of $ \Delta {\bf r}(t)$, but outline its solution. Let ${\bf r}_{phys}$ denote the solution in the physical frame that satisfies the original set and ${\bf r}_{LPT}$ denote the solution in the calculational frame obtained by the perturbative treatment discussed in the previous section. The two are related as   
\beq 
 {\bf r} = {\bf r}_{LPT} + \Delta {\bf r}(t).
 \label{frameshift}
 \eeq 
By substituting \eqnref{frameshift} in \eqnref{curleq} (with ${\bf v} =d {\bf r}/dt$), one obtains a differential equation for $ \Delta {\bf r}(t)$, where the source terms are determined by the LPT solution. The initial conditions for $\Delta {\bf r}$ are specified by the transformation between the physical frame and the calculational frame at the initial time. In the simple case of the plane Couette flow, we can assume $\Delta {\bf r}(0) =0$ and $\Delta {\dot {\bf r}}(0) =0$. The net physical solution and velocity 
\bea
{\bf r}({\bf R}, t) &=& {\bf R} + {\bf p}^{(0)}({\bf R}, t)  + {\bf p}^{(1)}({\bf R}, t) + \Delta {\bf r}(t), \\
{\bf v}_L({\bf R}, t) & =& {\dot {\bf p}}^{(0)}({\bf R}, t)    + {\dot {\bf p}}^{(1)}({\bf R}, t) +  \Delta {\dot {\bf r}}(t). 
\eea
Note however that the ${\bf v}$ is known as a function of the Lagrangian coordinate. In order to obtain the Eulerian velocity ${\bf v}_E({\bf r}, t)$  one has to solve for the initial ${\bf R}$ of the fluid element which is located at ${\bf r}$ at time $t$ i.e. if ${\bf r} = {\mathcal F}({\bf R}, t)$ then the Eulerian velocity at the coordinate point ${\bf r}$ is 
\beq 
{\bf v}_E({\bf r}, t) = {\bf v}_L({\mathcal F}^{-1}({\bf r}, t),t).
\label{backtoeuler}
\eeq

\section{Conclusion}
\label{sec:discussion}
The main motivation behind this paper was to outline the formalism of Lagrangian perturbation theory, a technique successful in other branches of physics, and apply it to the problem of flow stability. LPT has been used to model non-linear growth in cosmology for over four decades. Early work was done in the 1970's by Zeldovich \cite{zeldovich_gravitational_1970} with linear order perturbation theory. In the 1990s, others including Buchert and collaborators \cite{ehlers_newtonian_1997} developed higher orders and recently Nadkarni-Ghosh and Chernoff \cite{nadkarni-ghosh_extending_2011,nadkarni-ghosh_modelling_2013} addressed issues of convergence of the theory prior to orbit crossing. To the best of our knowledge, such techniques have been sparingly used to investigate the stability of fluid flows. As was outlined in the introduction, the fields of turbulence and large scale growth of structure in the universe share a common kind of complexity. In both, linear Eulerian stability theory is usually employed to get analytical results and the non-linear regime is often modeled using numerical simulations. It has often been the case that techniques developed in one branch of science later found utility in other seemingly unrelated branches of science. New techniques bring new insights and new ideas come from mapping concepts of one set of things to another. Higher order perturbation theory is always complicated, whether in Eulerian or Lagrangian frames, but is nevertheless useful to give analytical insights and can sometimes be used in conjunction with simulations to improve their efficiency. 

Here, we presented a first step in this direction. We focussed on the simplest shear flow: the plane Couette flow and restricted to 2D perturbations. This greatly simplified the expressions. In the Eulerian analysis, it is often enough to consider 2D perturbations thanks to Squire's theorem \cite{squire_1933}, which states that an unstable 3D eigenmode for some Reynolds number implies a unstable 2D eigenmode for a lower Reynolds number. Unfortunately, Squire's theorem, which is based on the Orr-Sommerfeld analysis, need not apply to the Lagrangian analysis so one may not be able to make conclusions based on 2D perturbations. Furthermore, it is known that transient growth is usually weaker in 2D than in 3D perturbations \cite{trefethen_hydrodynamic_1993}. For 3D perturbations or other types of flows the framework remains the same; of course the form of the equations is more complicated and their analysis will require numerics. This complexity is to be expected, but is not formally prohibitive. 

It is also possible to extend to the perturbative formalism to the non-linear regime by keeping terms to higher order in the displacement field in \eqnref{EXP}. Alternatively, it is possible to model the non-linear regime by repeated expansions of the linear PT (Nadkarni-Ghosh and Chernoff \mbox{\cite{nadkarni-ghosh_extending_2011,nadkarni-ghosh_modelling_2013})}.  This technique was initially developed in order to overcome the fact that, independent of orbit crossing, the Lagrangian series solution has a finite time range of validity. The basic idea is that LPT can be thought of as a numerical finite difference scheme with an associated time stepping criterion. Recent work \cite{zheligovsky_timeanalycity_2013, rampf_2015} also addresses the issue of analyticity of Lagrangian particle trajectories from a more formal perspective. The level of complexity of higher order LPT is perhaps comparable to higher order perturbation theory in the Eulerian frame. Each mode may have an associated time scale of evolution and their interaction may possibly complicate convergence. This phenomenon has been demonstrated on a two time-scale problem, where the scales are well separated: one fast and one slow (see Berry and Shukla \cite{berryshukla_2010} and references therein). It may be possible to apply LPT recursively to get the non-linear solution of the Navier-Stokes equation, however, issues of convergence, numerical stability etc. will need to be considered carefully to get meaningful results. Nevertheless, Lagrangian perturbative methods provide an alternate way to analyze the stability of fluid flows and the solutions could either be used in isolation or could be potentially useful in making educated guesses for starting numerical simulations that aim to understand the transition to turbulence.

\section{Appendix}

\subsection{Mathematical Transformations} 
\label{mathtrans}
We start with the expression for divergence of the velocity. 
\beq
\nabla_r\cdot {\dot {\bf r}} = \frac{\partial {\dot r}_i}{\partial r_i} = \frac{\partial {\dot r_i}}{\partial R_l}\frac{\partial R_l}{\partial r_i}.
\eeq
Einstein's repeated summation convention is followed. The inverse transformation from $R$-space to $r$-space is given as
\beq
\label{Rtor}
\frac{\partial R_l}{\partial r_i}= \frac{1}{2 J} \epsilon_{ilm}\epsilon_{jl'm'}\frac{\partial r_l}
{\partial R_{l'}}\frac{\partial r_m}{\partial R_{m'}},
\eeq
where
\beq
J={\rm Det}\left(\frac{\partial r_i}{\partial R_j}\right)=\epsilon_{jlm}\frac{\partial {r_1}}{\partial R_j}\frac{\partial {r_2}}{\partial R_l}\frac{\partial {r_3}}{\partial R_m}=\frac{1}{6}\epsilon_{ipq}\epsilon_{jlm}\frac{\partial  r_i}{\partial  R_j}\frac{\partial {r_p}}{\partial R_l}\frac{\partial { r_q}}{\partial R_m}
\label{jacobdef}
\eeq
and $\epsilon_{jlm}$ is the usual Levi-Civita symbol. The incompressibility condition reduces to 
\beq 
\epsilon_{ilm} \epsilon_{jl'm'} {\dot r}_{i,j} r_{l,l'} r_{m,m'} =0,  
\label{diveq2}
\eeq
where commas denote spatial derivatives with respect to the Lagrangian coordinate. For example, $r_{m,m'}$ denotes the derivative of the $m$-th component of the vector ${\bf r}$ with respect to the $m'$-th component of the Lagrangian coordinate ${\bf R} = \{ X,Y,Z\}$. Note that this can also be written as 
\beq
\epsilon_{ilm} \epsilon_{jl'm'} {\dot r}_{i,j} r_{l,l'} r_{m,m'} =\frac{1}{3} \epsilon_{ilm} \epsilon_{jl'm'}  \frac{d}{dt} \left(r_{i,j} r_{l,l'} r_{m,m'} \right).
\label{diveqaltform}
\eeq
Consider the curl of the Navier-Stokes equation. The $i$-th component of the l.h.s. is 
\bea 
\nabla_r \times {\ddot {\bf r}} &=& \epsilon_{ijk} \frac{\partial {\dot r}_k}{\partial r_j} \\ 
\nonumber        &=&   \epsilon_{ijk}  \frac{\partial {\dot r}_k}{\partial R_l} \frac{\partial R_l}{\partial r_j}\\
\nonumber        & =&  \frac{1}{2J} \epsilon_{ijk} \epsilon_{jmn} \epsilon_{lm'n'}  {\ddot r}_{k,l}r_{m,m'} r_{n,n'},
\eea
where the last equality follows from \eqnref{Rtor}. 
The r.h.s. of the Navier-Stokes is proportional to $\nabla_r^2(\nabla_r \times {\dot {\bf r}})$. For any scalar $f_i$, $\nabla_r$ converted to Lagrangian coordinates is 
\bea
\nabla_r^2 f_i &=& \frac{\partial}{\partial r_l}\left( \frac{\partial f_i}{\partial r_l} \right)\\
\nonumber &=& \frac{\partial}{\partial R_q} \left( \frac{\partial f_i}{\partial R_p} \cdot \frac{\partial R_p} {\partial r_l } \right) \cdot \frac{\partial R_q}{\partial r_l}.
\eea
Using \eqnref{Rtor} gives 
 \beq 
 \nabla_r^2 f_i = \frac{1}{2J} \epsilon_{lde} \epsilon_{qd'e'} r_{d,d'} r_{e,e'}  \left( \frac{1}{2J} \epsilon_{lab} \epsilon_{pa'b'} r_{a,a'} r_{b,b'} f_{i,p} \right)_{,q}.
 \eeq 
 Substituting $f_i = \epsilon_{ijk} \frac{\partial {\dot r}_k}{\partial r_j}$ and again using \eqnref{Rtor} to transform derivatives gives 
 \begin{flushleft}
 \beq 
\nabla_r^2 (\nabla_r \times {\dot {\bf r}})_i = \frac{1}{2J} \epsilon_{ijk} \epsilon_{jfg} \epsilon_{mf'g'} \epsilon_{lde} \epsilon_{qd'e'} \epsilon_{lab} \epsilon_{pa'b'}  r_{d,d'} r_{e,e'}   \left\{ \frac{1}{2J}  r_{a,a'} r_{b,b'} \left( \frac{1}{2J}  {\dot r}_{k,m} r_{f,f'} r_{g,g'}  \right)_{,p} \right\}_{,q}.
 \eeq
\end{flushleft}
 Thus the curl of the Navier-Stokes equation in Lagrangian coordinates reduces to 

\begin{flushleft}
\beq
\epsilon_{ijk} \epsilon_{jmn} \epsilon_{lm'n'}  {\ddot r}_{k,l}r_{m,m'} r_{n,n'} = \nu \epsilon_{ijk} \epsilon_{jfg} \epsilon_{mf'g'} \epsilon_{lde} \epsilon_{qd'e'} \epsilon_{lab} \epsilon_{pa'b'} 
r_{d,d'} r_{e,e'}   \left\{ \frac{1}{2J}  r_{a,a'} r_{b,b'} \left( \frac{1}{2J}  {\dot r}_{k,m} r_{f,f'} r_{g,g'}  \right)_{,p} \right\}_{,q}
\label{curleq2}
\eeq
\end{flushleft}

\subsection{The background solution in Lagrangian coordinates}
\label{zeroorder}
The solution for the physical position ${\bf r} = \{x,y,z\}$ in terms of the Lagrangian variable ${\bf R} = \{X,Y,Z\}$ at the zeroth order 
\beq 
{\bf r}({\bf R}) = {\bf p}^{(0)}({\bf R})= \{ X + cY t, Y,Z\}.
\eeq 
We check that this is an exact solution of the incompressible Navier-Stokes system given by \eqnrefs{mainT} and \eqnrefbare{mainL}. 
The transformation between the Lagrangian and Eulerian frame is 
\beq
\frac{\partial r_i}{\partial R_j} = \left( \begin{array}{ccc}
1 & ct & 0 \\
0 & 1& 0\\
0 & 0 & 1 \end{array} \right)
\eeq
and the inverse $\frac{\partial R_i}{\partial r_j}$ is given by \eqnref{Rtor} (or can be easily computed for this simple case), 
\beq 
\frac{\partial R_i}{\partial r_j} = \left( \begin{array}{ccc}
1 & -ct & 0 \\
0 & 1& 0\\
0 & 0 & 1 \end{array} \right).
\eeq
Here $i$ is the row-wise index, $j$ is column-wise. To check if \eqnref{mainT} and \eqnrefbare{mainL}, it suffices only to consider the $x$-component since the others are trivially zero. The divergence-less condition given by \eqnref{mainL} is 
\bea
\nabla_r{\dot  r}_x &=& 
 \frac{\partial {\dot r}_x}{\partial x}  \\
\nonumber &=& \frac{\partial {\dot r}_x}{\partial X } \cdot \frac{\partial X}{\partial x } + \frac{\partial {\dot r}_x}{\partial Y } \cdot \frac{\partial Y}{\partial x }\\
\nonumber &=& 0 \cdot 1 + c \cdot 0 =0
\eea

In \eqnref{mainT}, the l.h.s. is zero since there is no $t$ dependence in ${\dot {\bf r}}$ and Lagrangian derivative is just the total time derivative acting on ${\dot {\bf r}}$. So it remains to prove that r.h.s =0. The $x$-component is 
\bea
\nabla_r^2 {\dot {\bf r}} &=& \frac{\partial^2 {\dot r}_x}{\partial x^2} + \frac{\partial^2 {\dot r}_x}{\partial y^2}\\
\nonumber &=& \frac{\partial}{\partial x} \left(\frac{\partial {\dot r}_x}{\partial X}\cdot \frac{\partial X}{\partial x}  +\frac{\partial {\dot r}_x}{\partial Y} \cdot \frac{\partial Y}{\partial x} \right) + \frac{\partial}{\partial y} \left(\frac{\partial {\dot r}_x}{\partial X}\cdot \frac{\partial X}{\partial y}  +\frac{\partial {\dot r}_x}{\partial Y} \cdot \frac{\partial Y}{\partial y} \right).
\eea
Applying the change of derivatives again and using the fact that $\partial Y/\partial x =0$, $\partial {\dot r}_x/\partial X=0$ and 
$\partial^2 {\dot r}_x/\partial Y^2=0$, all terms become zero. Thus both Navier-Stokes and the incompressibility are satisfied when the base flow is expressed in Lagrangian coordinates. It is a natural candidate for the zeroth order particle position and one can examine the stability of the system by perturbing about this steady state solution.

\subsection{First order perturbation Theory} 
\label{perttheo}
The perturbation ansatz is ${\bf r} = {\bf p}^{(0)} +  {\bf p}^{(1)}\Delta$, where $\Delta$ is just a book-keeping parameter. Substitute this ansatz in \eqnrefs{diveq2} and \eqnrefbare{curleq2} and collect terms of first order. The zeroth order solution is determined by the background flow.  We will assume that the first order perturbation is two dimensional.

\subsubsection{Divergence Equation}
At first order \eqnref{diveqaltform} reduces to 
\beq
\epsilon_{ilm} \epsilon_{jl'm'} \frac{d}{dt} \left(p^{(0)}_{i,j} p^{(0)}_{l,l'} p^{(1)}_{m,m'}\right) \Delta =0.
\eeq
Using the symmetry properties of the Levi-Civita tensor and the fact that the background flow is laminar gives
\beq
\nabla_R \cdot {\dot {\bf p}^{(1)}} -{\dot p}^{(0)}_{X,Y} p^{(1)}_{Y,X} -  p^{(0)}_{X,Y} {\dot p}^{(1)}_{Y,X} =0.
\label{divsimp}
\eeq
Substituting for the plane Couette flow zeroth order solution from \eqnref{p0soln},
\beq 
\nabla_R \cdot {\dot {\bf p}^{(1)}} = c \left(p^{(1)}_{Y,X} + {\dot p}^{(1)}_{Y,X} t\right). .
\eeq
Note that a flow which is divergence-free in the Eulerian frame does not stay divergence-free in the Lagrangian frame. 

\subsubsection{Curl Equation}
To simplify the curl equation we first note that from \eqnref{jacobdef} the determinant $J$ to first order can be expanded as 
\beq
J = \frac{1}{6} \epsilon_{ipq}\epsilon_{jlm} \left(\frac{\partial  p^{(0)}_i}{\partial  R_j}\frac{\partial {p^{(0)}_p}}{\partial R_l}\frac{\partial { p^{(0)}_q}}{\partial R_m} + 3\frac{\partial  p^{(0)}_i}{\partial  R_j}\frac{\partial {p^{(0)}_p}}{\partial R_l}\frac{\partial { p^{(1)}_q}}{\partial R_m} \cdot \Delta \right) + \mathcal{O}(\Delta^2).
\eeq
When the background flow is laminar this gives 
\beq 
J = 1+  \left(\nabla_R \cdot {\bf p}^{(1)} - p^{(0)}_{X,Y} p^{(1)}_{Y,X} \right) \Delta+\mathcal{O}(\Delta^2).
\label{firstorderJ}
\eeq
But note that \eqnref{divsimp} can be re-written as 
\beq 
\frac{d}{dt} [\nabla_R \cdot {\bf p}^{(1)} - p^{(0)}_{X,Y} p^{(1)}_{Y,X}] =0.
\eeq
Comparing with the expression \eqnref{firstorderJ} for $J$, this gives ${\dot J} =0$ to first order or in other words $J$ is conserved to first order. 
Since $J(t=0) =1$, by definition of the Lagrangian coordinate, to first order we can set $J \approx 1$ in \eqnref{curleq2}. The coefficient of the first order term in $\Delta$ in the l.h.s. of \eqnref{curleq2} is 
\beq 
 2  \left({\ddot p}^{(1)}_{Y,X} - {\ddot p}^{(1)}_{X,Y}  \right) + 2 p^{(0)}_{X,Y} {\ddot p}^{(1)}_{X,X}.
\label{lhscurl}
\eeq
We now consider the structure of the r.h.s. of \eqnref{curleq2}. We remind the reader that the 'comma' subscript denotes differentiation w.r.t. the Lagrangian coordinate. First note that each $r$ will have an expansion in terms of $p^{(0)}$ and $p^{(1)}$ and, at first order, any term involving $p^{(1)}$ can only couple to other terms with $p^{(0)}$. Depending on the location of $p^{(1)}$, we can classify these into four types of terms. The first type is when $p^{(1)}$ is located between the curly and regular bracket, the second is when $p^{(1)}$ is within the regular bracket and with the time derivative, the third is when it is within the regular bracket but the time derivative does not act on it and the fourth is when it is outside both brackets. Thus, the coefficient of the term which is first order in $\Delta$ in the r.h.s. of \eqnref{curleq2} is I + II + III + IV, with, 
\bea 
{\rm I.} \; \; \;  2 \mathcal{E}  && \frac{\partial}{\partial R_q} \left\{p^{(1)}_{a,a'} p^{(0)}_{b,b'}\frac{\partial}{\partial R_p} \left( {\dot p}^{(0)}_{k,m},p^{(0)}_{f,f'},p^{(0)}_{g,g'} \right) \right\} p^{(0)}_{d,d'} p^{(0)}_{e,e'}\\
{\rm II.} \; \; \; \; \;   \mathcal{E}  && \frac{\partial}{\partial R_q} \left\{p^{(0)}_{a,a'} p^{(0)}_{b,b'}\frac{\partial}{\partial R_p} \left( {\dot p}^{(1)}_{k,m},p^{(0)}_{f,f'},p^{(0)}_{g,g'} \right) \right\} p^{(0)}_{d,d'} p^{(0)}_{e,e'}\\
{\rm III.} \; \; \; 2\mathcal{E}  && \frac{\partial}{\partial R_q} \left\{p^{(0)}_{a,a'} p^{(0)}_{b,b'}\frac{\partial}{\partial R_p} \left( {\dot p}^{(0)}_{k,m},p^{(0)}_{f,f'},p^{(1)}_{g,g'} \right) \right\} p^{(0)}_{d,d'} p^{(0)}_{e,e'}\\
{\rm IV.} \; \; \; 2 \mathcal{E}  && \frac{\partial}{\partial R_q} \left\{p^{(0)}_{a,a'} p^{(0)}_{b,b'}\frac{\partial}{\partial R_p} \left( {\dot p}^{(0)}_{k,m},p^{(0)}_{f,f'},p^{(0)}_{g,g'}\right) \right\} p^{(0)}_{d,d'} p^{(1)}_{e,e'}
\eea
where $\mathcal{E} = \frac{\nu}{4} \epsilon_{ijk} \epsilon_{jfg} \epsilon_{mf'g'} \epsilon_{lde} \epsilon_{qd'e'} \epsilon_{lab} \epsilon_{pa'b'}$. 
For the plane Couette flow terms of the type I and IV will be zero since there are two spatial derivatives acting on components of ${\bf p}^{(0)}$. 

The terms of type II and III simplify to 
\bea
{\rm II} \rightarrow&& \nu \left(2 \nabla_R^2 -4 ct \frac{\partial^2}{\partial X \partial Y } + 2 c^2 t^2 \frac{\partial}{\partial X^2}  \right)
 \left({\dot p}^{(1)}_{Y,X} - {\dot p}^{(1)}_{X,Y}  +ct {\dot p}_{X,X} ^{(1)}\right).
\label{rhscurl1}
\\
{\rm III} \rightarrow&& \nu \left(2 \nabla_R^2 -4 ct \frac{\partial^2}{\partial X \partial Y } + 2 c^2 t^2 \frac{\partial}{\partial X^2}  \right) (-c p^{(1)}_{X,X}).
\label{rhscurl2}
\eea
 Putting together \eqnrefs{lhscurl} and \eqnrefbare{rhscurl1} and \eqnrefbare{rhscurl2} gives 

\begin{flushleft}
  \beq
   \left({\ddot p}^{(1)}_{Y,X} - {\ddot p}^{(1)}_{X,Y}\right) +  ct {\ddot p}^{(1)}_{X,X} = \nu \left( \nabla_R^2 -2 ct \frac{\partial^2}{\partial X \partial Y } +  c^2 t^2 \frac{\partial}{\partial X^2}  \right) \left(  {\dot p}^{(1)}_{Y,X} - {\dot p}^{(1)}_{X,Y}  +ct {\dot p}_{X,X} ^{(1)} -c p^{(1)}_{X,X}\right).
 \eeq
 \end{flushleft}

\subsection{Conditions for the bounded Couette flow}
\label{app:bounded}
For the bounded Couette flow, we can take $\phi$ to be of the general form 
\beq 
\phi (X,Y,t) = H(X,Y) f(t), 
\label{phiform2}
\eeq
where $H(X,Y)$ will be an appropriately chosen `basis function' which satisfies the boundary conditions at both plates. 
Substituting it in \ref{simpleeq2} we get 
\beq 
\frac{d f}{dt} = \nu \left((1+c^2t^2)\frac{H,_{XX}}{H} +\frac{H,_{YY}}{H} - 2 ct  \frac{H,_{XY}}{H} \right) f. 
\eeq
Here $H,_{XY} = \frac{\partial^2 H}{\partial X \partial Y}$ etc. Since $H$ is just a function of space, this can be integrated w.r.t. time to give 
\beq 
f(t) = f(0) \exp{
\left[ \nu t \left(\frac{H,_{XX} + H,_{YY}}{H}\right)  + \frac{ \nu c^2 t^3}{3} \frac{H,_{XX}}{H} -  \nu ct^2 \frac{H,_{XY}}{H} \right] }
\eeq
where $f(0)$ is the integration constant to be set later. One can now solve \eqnref{simpleeq1} for ${\dot \psi}$ with the ansatz 
\beq 
{\dot \psi}(X,Y,t) = G(X,Y,t) f(t). 
\eeq
Here $\phi$ acts like a source term so the temporal dependence of $G(X, Y, t)$ can always be chosen to be of the above form. Substituting in \eqnref{simpleeq1} and using the form of $\phi$ gives the equation 
\beq 
(1+c^2 t^2) G,_{XX} + G,_{YY} -2ct G,_{XY} = H(X,Y). 
\label{Geq}
\eeq 
Given a $H(X,Y)$, this equation is a PDE in two variables which needs to be solved numerically subject to the boundary conditions. 

For infinite extent along $x$ direction, it is possible to use Fourier decomposition along the $x$-axis: $H(X,Y) = \sum_{k_x} \tilde{\phi}(k_x) e^{ik_x X} h(Y)$ and $G(X,Y,T) = \sum_{k_x} \tilde{\phi}(k_x) e^{ik_x X} g(Y,t)$. \capeqnref{Geq} becomes, 
\beq
-k_x^2 (1+c^2 t^2) g(Y,t) -2 ik_x ct g'(Y,t) + g''(Y,t) = h(Y); 
\eeq 
where the primes denote differentiation w.r.t $Y$. 
The solution for $g(Y,t)$ can be split into a homogeneous part and a particular solution: $g(Y,t) = g_{homo.}(Y,t) + g_{part.}(Y,t)$. The homogeneous solution is 
\beq 
g_{homo.}(Y,t) = C_1(t) g_1(Y) + C_2(t) g_2(Y)
\eeq
where  $g_1(Y) = e^{m_1Y}$ and $g_2 = e^{m_2 Y}$ with $m_1 = -k_x + ik_x ct $ and $m_2 = k_x + ik_x ct$. The particular solution is given by 
\beq 
g_{part.}(Y,t) = a(Y) g_1(Y) + b(Y) g_2(Y)  
\eeq
where 
\beq 
a'(Y) = -\frac{h(y)}{W(g_1,g_2)} g_2(Y) \;\;\;\; b'(Y) = \frac{h(y)}{W(g_1,g_2)} g_1(Y), 
\eeq
where the Wronskian $W(g_1,g_2) = g_1 g_2'-g_1'g_2$. In this case the two linearly independent homogeneous solutions are exponentials and $W(g_1,g_2) = (m_2-m_1) g_1 g_2$. This gives the full solution as 
\beq
g(Y,t) = c_1 g_1(Y) + c_2 g_2(Y) -\frac{g_1(Y)}{m_2-m_1} \int_0^Y \frac{h(Y')}{g_1(Y')} dY' + \frac{g_2(Y)}{m_2-m_1} \int_0^Y \frac{h(Y')}{g_2(Y')} dY
\eeq
The boundary conditions given by \eqnrefs{bc1} and \eqnrefbare{bc2}  get extended to 
\beq
\left. \frac{\partial {\dot \psi}}{\partial X}\right|_{Y=0, h}=0 \;\;\; {\rm and } \;\;\; \left. \frac{\partial {\dot \psi}}{\partial Y}\right|_{Y=0,h} = 0.\; \; \; \; \; \; \forall \; t.
\eeq
This imposes constraints on $c_1, c_2$ and the form of $h$. The important point to note here is that the temporal dependence for the bounded flow has also the same exponential factor as the semi-bounded case and it is plausible that it will exhibit the same late time behaviour: the flow will be stable for all Reynolds numbers (or all values of kinematic viscosity). 

\subsection{Effect of the $({\bf v} \cdot \nabla) {\bf v} $ term in Eulerian theory}
\label{app:trans}
Let ${\bf v}_{s.s} = \{cy,0\}$ denote the steady state solution for the plane Couette flow and let ${\bf v}$ denote the perturbation around this background solution. The full non-linear equation satisfied by ${\bf v}$ is 
\beq 
\frac{\partial {\bf v}}{\partial t} =-({\bf v}_{s.s} \cdot \nabla) {\bf v} -({\bf v} \cdot \nabla) {\bf v}_{s.s} + \nu \nabla^2 {\bf v} + ({\bf v} \cdot \nabla){\bf v},
\eeq
where the spatial derivatives are w.r.t. the Eulerian coordinate ${\bf r}\equiv \{x,y\}$. 
This can be written as 
\beq
\frac{\partial {\bf v}}{\partial t} = {\mathcal L} \cdot {\bf v}+  ({\bf v} \cdot \nabla) {\bf v}, 
\eeq
 where the operator ${\mathcal L}$ for the plane Couette flow is, 
 \beq 
 {\mathcal L} = \left( \begin{array}{cc}
 -c y \frac{\partial}{\partial x} + \nu \nabla^2& -c \\
0 & -cy  \frac{\partial}{\partial x} + \nu \nabla^2 
 \end{array} \right).
\eeq
 We construct the solution to the full non-linear solution iteratively as follows. 
 Let ${\bf v}^{(1)}$ be the linear solution that satisfies 
 \beq
\frac{\partial {\bf v}^{(1)}}{\partial t} = {\mathcal L} \cdot {\bf v}^{(1)}.
\eeq
Integrating over $t$ for fixed $x,y$, we get 
 \beq 
 {\bf v}^{(1)}(x,y,t) =  e^{{\mathcal L}t} \cdot {\bf v}_0,
 \eeq
where ${\bf v}_0$ is the perturbation at the initial time $t=0$. We assume it to have the form 
 \beq {\bf v}_{0}(x,y)= \{v_{x,0} ,v_{y,0}\} e^{i {\bf k} \cdot {\bf r}}. \eeq
 Use this solution to construct the non-linear acceleration as $({\bf v}^{(1)} \cdot \nabla) {\bf v}^{(1)}$. For early times, the second order solution is 
 \beq 
 {\bf v}^{(2)}(x,y,t) = {\bf v}^{(1)}(x,y,t) + [({\bf v}^{(1)} \cdot \nabla) {\bf v}^{(1)}] \cdot t.
  \eeq
  
\begin{figure}
 \includegraphics[width=18cm]{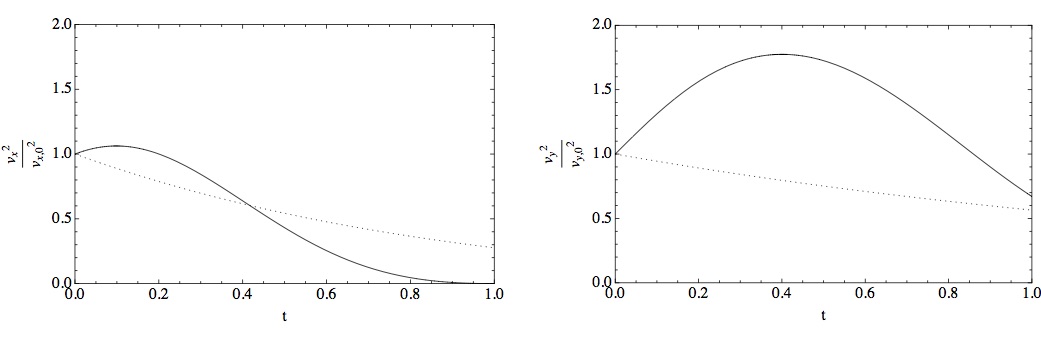}
 \caption{The ratio of the linear (dotted lines) and second order (solid lines) Eulerian perturbation solutions at early times. The left and right panels correspond to the $x$ and $y$-components respectively. Note that the second order solution which is obtained using by including the convective term $({\bf v} \cdot \nabla) {\bf v}$ (constructed from the first order solution) shows transient growth before eventually decaying away. }
 \label{transfig}
 \end{figure}

It can be seen from the form of the operator ${\mathcal L}$ that all modes will be stable at linear order. The form of the linear operator is such that there is only one Eigen direction along $(1,0)$ and the linear mechanism of non-normal growth which relies on the non-orthogonality of the eigenvectors \cite{schmidbook} is not applicable here.  
However, for some modes, the second order solution may show hints of transient growth. We evaluate numerically $ {\bf v}^{(1)}$ and $ {\bf v}^{(2)}$ with parameters $\nu =0.1, c=1, x=1,y=1,kx=1,ky=1$ (this choice allows us to demonstrate the transient growth). \capfigref{transfig} shows the plots of the ratio of final velocity to initial velocity for small time $t < c^{-1}$. The dotted (solid) line denotes the first (second) order solution and the left (right) panel is the $x$ ($y$) component. 
 It is clear that the linear solution ${\bf v}^{(1)}$ decays exponentially, whereas the second order solution that is calculated using the term $({\bf v} \cdot \nabla) {\bf v}$ shows a growth at early times. Thus, in this case (plane Couette) the transient growth is not seen at linear order but is hinted at second order. The main point here is that the second order Eulerian solution obtained perturbatively is in qualitative agreement with the linear Lagrangian solution. As mentioned in the text, matching the two frames orderwise is ill-defined. A true comparison of the solutions can be made only in the same frame (Eulerian) and is beyond the scope of the current work.

\section*{Conflict of Interest Statement}
The authors declare that the research was conducted in the absence of any commercial or financial relationships that could be construed as a potential conflict of interest.

\section*{Author Contributions}
S. N-G. and J.K. B. formulated the problem. S. N-G. performed the analytic calculations and wrote the paper. J.K.B. supervised the project. 

\section*{Acknowledgments}
S. N-G. would like to thank Mahendra Verma, Stefano Anselmi and Rama Govindrajan for valuable discussions and suggesting useful references. Thanks are due also to Sagar Chakraborty for discussions and comments on the manuscript. 

\section*{Funding}
S.N-G. would like to thank the Science and Engineering Research Board (SERB) of the Department of Science and Technology (DST), Govt. of India for the grant YSS/2014/000526.

\end{document}